\documentclass[
	aps,
	groupeaddress,
	longbibliography,
]{revtex4-2}

\usepackage[english]{babel}
\usepackage[utf8]{inputenc}
\usepackage{amsmath,amssymb,amsfonts}
\usepackage{color}
\usepackage{graphicx}
\usepackage{physics}
\usepackage[normalem]{ulem}
\usepackage{tabularx}
\usepackage{siunitx}

\newcommand{\change}{}

\renewcommand{\selectlanguage}[1]{} 

\usepackage[
	colorlinks=true,
	linkcolor=blue,
	urlcolor=blue,
	citecolor=blue
]{hyperref}

\begin{document}

\title{Theory of single-photon emission from neutral and charged excitons in a polarization-selective cavity}

\newcommand{\dtu}{DTU Electro, Department of Electrical and Photonics Engineering, Technical University of Denmark, 2800 Kongens Lyngby, Denmark}

\author{Luca Vannucci}
\email{lucav@dtu.dk}
\affiliation{\dtu}

\author{Niels Gregersen}
\affiliation{\dtu}

\date{\today}

\begin{abstract}

Single-photon sources based on neutral or charged excitons in a semiconductor quantum dot are attractive resources for photonic quantum computers and simulators.
To obtain indistinguishable photons, the source is pumped on resonance with polarized laser pulses, and the output is collected in orthogonal polarization.
However, for sources featuring vertical emission of light, 50\% of the emitted photons are unavoidably lost in this way.
Here, we theoretically study the quantum dynamics of an exciton embedded in an asymmetric vertical cavity that favors emission in a specific polarization. We identify the configuration for optimal state initialization and demonstrate a path toward near-unity polarized efficiency.
We also derive simple analytical formulas for the photon output in each polarization as a function of the Purcell-enhanced emission rates, which shed light on the physical mechanism behind our results.

\end{abstract}

\maketitle

\section{Introduction}

Highly efficient emitters of indistinguishable single photons are essential resources for photonic quantum computers and simulators, where one qubit of information is encoded in the quantum state of a single photon \cite{Maring2024}.
A calculation involving $N$ qubits requires $N$ single photons as the input state. Given a source with efficiency $\varepsilon$, the success probability for generating the required input scales exponentially as $\varepsilon^N$, showing that $\varepsilon$ must be as close to 1 as possible.
Moreover, it is required that the emitted photons are indistinguishable from each other for a successful quantum computation \cite{Aaronson2011}.

Spontaneous parameteric downconversion is a popular and straightforward technique to generate highly indistinguishable heralded photons \cite{Kwiat1995}. However, it is a probabilistic process with limited single-photon purity and collection efficiency, which requires complex multiplexing schemes \cite{Joshi2018, Kaneda2019}.
As alternative, a well-established and 
deterministic strategy to produce single photons on demand is to place a semiconductor quantum dot (QD) inside an optical cavity, whose role is to enhance the emission rate and route the photons towards the collection setup \cite{Heindel2023}.
For example, a QD inside an optimized micropillar cavity is predicted to reach an efficiency of $\varepsilon = 0.95$ in theory \cite{Wang2020_PRB_Biying}.

However, it is well known that the state preparation protocol has a significant influence on the performance.
For example, above-band excitation of the QD is straightforward to implement and allows spectral filtering of the pumping laser; however, it leads to poor indistinguishability due to significant time-jitter and charge noise \cite{Huber2015}.
The best indistinguishability is obtained under pulsed resonant excitation, whereby the pumping laser is tuned to resonance with the natural frequency of the emitter, and has therefore the same frequency as the outgoing single-photons \cite{Somaschi2016}.
For sources based on photonic crystal cavities, it is possible to separate the incoming laser light from the desired single-photon without introducing additional losses \cite{Uppu2020, Zhou2022}.
On the other hand, for structures with vertical emission such as the micropillar \cite{Wang2020_PRB_Biying, Somaschi2016}, the laser is rejected via a cross-polarization setup, which reduces the total system efficiency.
Ollivier \textit{et al} \cite{Ollivier2020} have carefully analyzed the dynamics in this scenario, considering a neutral exciton with two orthogonal dipole states embedded in a circular micropillar cavity. They have shown that the best configuration is obtained when the pumping laser is polarized at \SI{45}{\degree} with respect to the exciton axes. Under this condition, and assuming ideal source efficiency $\varepsilon = 1$, the fraction of photons collected in the orthogonal polarization is $\eta = \frac 1 2 \frac{\Delta_{\rm FSS}^2}{\Gamma^2 + \Delta_{\rm FSS}^2}$, where $\hbar \Delta_{\rm FSS}$ is the fine structure splitting (FSS) between the exciton states, and $\Gamma$ is the spontaneous emission rate. Clearly, $\eta$ is fundamentally limited to 0.50, which is well below the requirements for scalable quantum computation involving many qubits.
Considering instead a charged exciton (or trion), there are no requirements in terms of polarization alignment and FSS. However, $\eta$ is still limited to 0.50 due to equal emission into orthogonal polarizations.

It is thus natural to ask how to break the 0.50 threshold while still resorting to resonant excitation to ensure optimal indistinguishability.
To increase the emission rate in a specific polarization, Wang \textit{et al} \cite{Wang2019a} fabricated asymmetric cavities with an elliptical shape rather than circular, reporting an efficiency of 0.60.
A similar strategy underlies the open cavity approach, where polarization-dependent coupling is made possible by birefringence of the host semiconductor \cite{Tomm2021, Ding2025}, and source efficiency of 0.71 has recently been demonstrated \cite{Ding2025}.
Theoretical results have shown that an optimized elliptical micropillar could reach an efficiency of 0.90 \cite{Gür2021}. This has stimulated further investigations of elliptical micropillars \cite{Kazanov2023} and Bragg gratings \cite{Ge2024, Chen2022}, with a focus on improving the optical design.
However, the quantum dynamics of an exciton embedded in an elliptical cavity, which fundamentally determines the performance of these devices, has not been studied so far.

In this work, we fill this gap and report the theoretical study of the dynamics of an exciton subjected to polarization-selective spontaneous emission.
Considering first the case of a neutral exciton, we present a model of a three-level quantum emitter embedded in a cavity supporting two non-degenerate and orthogonally-polarized modes, and consider arbitrary rotation between exciton and cavity polarization axes.
Solving the master equation numerically, we demonstrate a path to improve the polarized efficiency arbitrarily close to unity under realistic experimental conditions. We discuss the role of the FSS and the relative alignment between the cavity and exciton axes, and show that the conditions described in Ref.~\cite{Ollivier2020} to optimize the performance of circularly-symmetric structures (i.e.\ \SI{45}{\degree} between cavity and exciton axes, and nonzero FSS) hold also for the case of asymmetric devices.
Moreover, we solve the model analytically in the weak-coupling regime of cavity quantum electrodynamics. We provide a simple analytical expression for the number of photons emitted per excitation pulse in each polarization, as a function of the polarization-selective emission rates and of the FSS.
Then, we extend the formalism to the case of a charged exciton and discuss its performance in comparison with a charge-neutral emitter.

Our results demonstrate that polarization-selective cavities are an ideal platform in the quest for single-photon sources (SPS) with near-unity efficiency and indistinguishability.
Our theory provides valuable support for optical simulations of SPS devices, where assumptions must be made about the power emitted in each polarization \cite{Gür2021, Wang2022_nanobeam}. Moreover, it provides guidelines for the fabrication of SPS based on elliptical cavities and will stimulate additional experimental work.

\section{Model --- Neutral exciton}
\label{sec:model}

\begin{figure}
    \centering
    \includegraphics[width=1.0\linewidth]{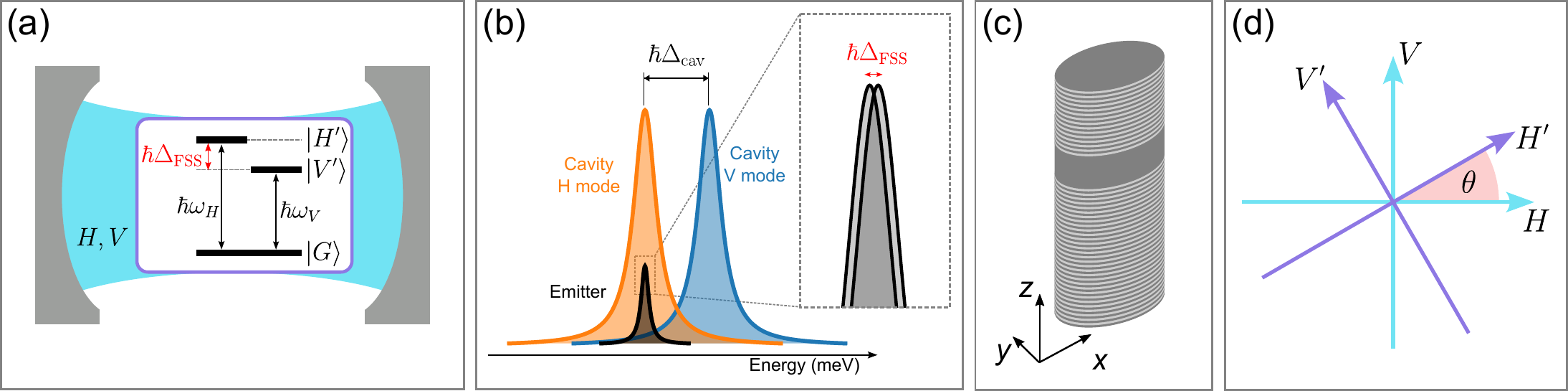}
    \caption{
    (a) Energy levels of a neutral exciton with eigenstates $\ket{G}, \ket{V'}, \ket{H'}$ and fine structure splitting $\hbar \Delta_{\rm FSS}$, embedded in a cavity supporting two non-degenerate modes $H$ and $V$.
    (b) Energy spectrum of the cavity modes and of the emitter, showing the splitting $\hbar \Delta_{\rm cav}$ between cavity modes. The inset shows the splitting $\hbar \Delta_{\rm FSS}$ between exciton states, which is much smaller than $\hbar \Delta_{\rm cav}$.
    (c) Sketch of an elliptical micropillar supporting two non-degenerate cavity modes.
    (d) Exciton axes are rotated by an angle $\theta$ with respect to the cavity axes.
    }
    \label{fig:fig1}
\end{figure}

We consider a neutral exciton with excited eigenstates $\ket{H'}$ and $\ket{V'}$ corresponding to orthogonal polarizations of the emitted photons, and ground state $\ket{G}$ sketched in Fig.~\ref{fig:fig1}(a). The Hamiltonian is $H_{\rm e} = \hbar \omega_H \dyad{H'} + \hbar \omega_V \dyad{V'}$, where the transition frequencies $\omega_H = \omega_0 + \frac 1 2 \Delta_{\rm FSS}$ and $\omega_V = \omega_0 - \frac 1 2 \Delta_{\rm FSS}$ differ by the fine structure splitting $\Delta_{\rm FSS} = \omega_H - \omega_V$.
The emitter is placed in a cavity supporting two non-degenerate modes with resonant frequencies $\omega_{{\rm cav}, j}$, $j \in \qty{H, V}$, and annihilation operators $a_j$. 
For each mode, the cavity frequency is related to $\omega_0$ via $\omega_{{\rm cav}, j} = \omega_0 + \delta_j$, with $\delta_j$ the mode-dependent cavity detuning.
As suggested in Fig.~\ref{fig:fig1}(b), we will assume that the $H$ cavity mode is on resonance with the exciton, while the $V$ mode is shifted away by the cavity splitting $\hbar \Delta_{\rm cav} = \hbar \delta_V - \hbar \delta_H$.
As a specific example, we will consider the case of an elliptical micropillar cavity [see Fig.~\ref{fig:fig1}(c)], where $\hbar \Delta_{\rm cav}$ can reach a few hundreds of \SI{}{\micro\eV} and is thus much larger than the FSS \cite{Wang2019a, Gür2021}.
Nonetheless, our theoretical formalism is valid for any cavity that supports energy splitting between modes with orthogonal polarization.
This could be due, for example, to the application of uniaxial stress \cite{Tomm2021_uniaxial_stress, Tomm2021, Ding2025} or to a small inherent anisotropy of the crystal lattice.

We consider an arbitrary rotation of angle $\theta$ between the cavity axes and the exciton eigenstates.
Therefore, it is convenient to define a rotated set $\qty{\ket{H}, \ket{V}}$, for the exciton, where the new basis states are aligned with the respective cavity polarizations [see Fig.~\ref{fig:fig1}(d)]. The rotated exciton states are
\begin{align}
    \ket{H} & = \cos(\theta) \ket{H'} - \sin(\theta) \ket{V'} ,\\
    \ket{V} & = \sin(\theta) \ket{H'} + \cos(\theta) \ket{V'} .
\end{align}
Moving to a frame rotating at frequency $\omega_0$, the system Hamiltonian in the rotated basis becomes
\begin{align}
\label{eq:model}
    H
    & = \frac{\hbar \Delta_{\rm FSS}}{2} \cos(2\theta) \qty(\dyad{H} - \dyad{V}) + \frac{\hbar \Delta_{\rm FSS}}{2} \sin(2\theta) \qty(\dyad{H}{V} + \dyad{V}{H}) \nonumber \\
    & \quad + \sum_{j = H, V} \qty[\hbar \delta_j a_j^\dag a_j + \hbar g_j \qty(\dyad{j}{G} a_j + \dyad{G}{j} a_j^\dag)] .
\end{align}
The last term in Eq.~\eqref{eq:model} accounts for cavity-exciton interaction via two copies of a Jaynes-Cummings Hamiltonian \cite{Jaynes1963}, one per each polarization. The parameter $g_j$ is the polarization-dependent light-matter coupling strength.
In the following, we assume that the polarizations of excitation and collection are aligned with the cavity axes $V$ and $H$, respectively \cite{Ollivier2020}.

We calculate the system dynamics with a master equation approach.
We assume lossy cavities with loss rates $\kappa_j$, and model photon losses with two Lindblad terms $\kappa_j \mathcal L_{a_j}[\rho]$, with $\mathcal L_A[\rho] = A \rho A^\dag - \frac{1}{2} \acomm{A^\dag A}{\rho}$.
Moreover, we assume that the exciton couples to other non-cavity background modes, and this is modeled with two additional terms $\Lambda_j \mathcal L_{\dyad{G}{j'}}[\rho]$, with $\Lambda_j$ the emission rate into non-cavity modes. 
The master equation is thus
\begin{align}
\label{eq:master}
    \dv{t} \rho
    = -\frac{i}{\hbar} \comm{H}{\rho} + \sum_{j = H, V} \kappa_j \mathcal L_{a_j} [\rho] + \sum_{j = H, V} \Lambda_j \mathcal L_{\dyad{G}{j'}} [\rho] .
\end{align}
To obtain $\rho(t)$, we represent operators as $3 \times 3$ matrices in the basis $\qty{\ket{H}, \ket{V}, \ket{G}}$ and solve Eq.~\eqref{eq:master} numerically with a 4th order Runge-Kutta solver for a given initial state $\rho(0)$, and up to the final time $t_{\rm f}$. We carefully check convergence with respect to the time step $\dd{t}$ and the final time $t_{\rm f}$. The latter must be chosen large enough so that full relaxation to the ground state is obtained, i.e.\ $\rho(t_{\rm f}) = \dyad{G}$.

We quantify the polarized source efficiency as the number of photon emitted from the cavity in each polarization,
\begin{equation}
\label{eq:Nj}
    N_j = \kappa_j \int_0^{+\infty} \dd t \Tr \qty[a_j^\dag a_j \rho(t)] .
\end{equation}
Correspondingly, the number of photons emitted into background modes is  
\begin{equation}
\label{eq:Bj}
    B_j = \Lambda_j \int_0^{+\infty} \dd t \Tr \qty[\dyad{j'} \rho(t)] ,
\end{equation}
and we assume that these photons are lost are and never detected.
Clearly, if the system is initialized in a single-excitation state, the condition $N_{\rm tot} = N_V + N_H + B_V + B_H = 1$ must be fulfilled for the total number of emitted photons.
It should be noted that this simplified picture, in which only the cavity photons contribute to the efficiency while the background emission is completely lost, is well suited when the emission can be explained in terms of a single optical mode.
This assumption is well justified for established designs such as the micropillar \cite{Wang2020_PRB_Biying} and the photonic nanowire \cite{Friedler2009, Claudon2010}, and similar definitions to Eqs.~\eqref{eq:Nj} and \eqref{eq:Bj} are commonly encountered in the literature \cite{Cosacchi2019, Gustin2020, Vannucci2023, Heinisch2024}.
However, the single-optical-mode approximation breaks down for smaller structures such as the nanopost \cite{Gaignard2025}, where background modes interfere with the fundamental cavity mode and give substantial contribution to the efficiency \cite{Jacobsen2023}.
Furthermore, we implicitly assume in Eq.~\eqref{eq:Nj} that every photon that leaks out of the cavity is eventually collected, which is unrealistic.
Out-coupling and the role of the collection setup can be modeled with an additional factor $\epsilon$ on the right hand side of Eq.~\eqref{eq:Nj}, with $0 \leq \epsilon \leq 1$. However, accurate modeling of optical collection goes beyond the scope of this work because it is not related to the quantum dynamics of the exciton state. Therefore, we assume $\epsilon=1$ for simplicity.

\section{Results}
\label{sec:results}

\subsection{Numerical solution of the master equation}
\label{sec:results_numerical}

We apply our formalism to the case of a neutral exciton emitting at \SI{900}{\nano\meter} embedded in a micropillar cavity. 
Unless stated otherwise, the model parameters are $\hbar g_H = \hbar g_V = $ \SI{26.3}{\micro\eV}, $\hbar \kappa_H = \hbar \kappa_V = $ \SI{329}{\micro\eV}, $\hbar \Lambda_H = \hbar \Lambda_V = $ \SI{0.66}{\micro\eV}. These values are the same as in Ref.~\cite{Vannucci2023} and correspond to a circular micropillar of diameter $D = $ \SI{1.84}{\micro\meter} with 15 layer pairs in the top DBR, leading to Q-factor of approximately 4500.
The coupling constants $g_j$, loss rates $\kappa_j$, and background emission rates $\Lambda_j$ are calculated from optical simulations of the electromagnetic environment \change{using the Fourier Modal Method \cite{Gur2021, Wang2020_PRB_Biying}.}

We assume that the $V$ cavity mode is used for excitation and the $H$ mode for collection.
We choose not to model the state preparation dynamics for simplicity, although this could be done by introducing an extra term in the Hamiltonian coupling a time-dependent laser pulse to the $V$ dipole (see e.g. Ref.~\cite{Piccinini2025_SUPER}).
Instead, we initialize the system in $\rho(0) = \dyad{V}$ at the initial time $t=0$, and the objective is to find a configuration that maximizes the number of photons $N_H$ emitted in the orthogonal polarization. 

\begin{figure*}
    \centering
    \includegraphics[width=\linewidth]{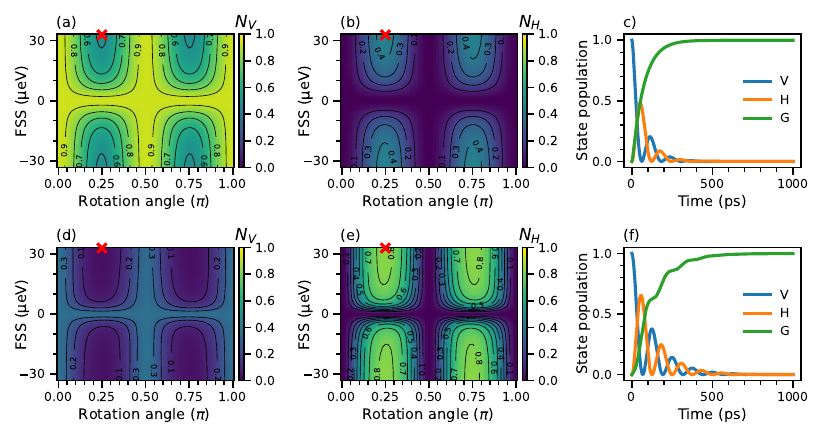}
    \caption{\textbf{(a--c)} Polarized photon emission and exciton dynamics for a symmetric structure with degenerate cavity modes, $\hbar \Delta_{\rm cav} = 0$.
    Panels (a) and (b) show the number of photons emitted with vertical ($N_V$) and horizontal ($N_H$) polarization as a function of the rotation angle $\theta$ between cavity and exciton axes and of the FSS $\hbar \Delta_{\rm FSS}$.
    Panel (c) show the evolution of the emitter state population in time for $\theta = \pi/4$ and $\hbar \Delta_{\rm FSS} = $ \SI{32.9}{\micro\eV}, corresponding to the red cross in (a, b).
    \textbf{(d--f)} Same as in (a-c), with nonzero splitting $\hbar \Delta_{\rm cav} = $ \SI{770}{\micro\eV} between the cavity modes.
    }
    \label{fig:fig2}
\end{figure*}

First, in Figs.~\ref{fig:fig2}(a) and \ref{fig:fig2}(b) we consider the case of degenerate cavity modes, $\hbar \Delta_{\rm cav} = 0$, which is suitable for a circularly symmetric micropillar \cite{Wang2020_PRB_Biying}.
We observe that $N_V$ is maximum and $N_H$ is exactly zero when the cavity axes are parallel to the exciton axes and when $\Delta_{\rm FSS} = 0$, while we find a significant reduction in $N_V$ and a simultaneous increase in $N_H$ for non-zero FSS and rotation angle $\theta$. The maximum value of $N_H$ approaches 0.5 and occurs at the largest value of FSS considered ($\hbar \Delta_{\rm FSS} =$ \SI{32.9}{\micro\eV}) and $\theta = \frac{\pi}{4}$. This configuration is marked with a red cross in Figs.~\ref{fig:fig2}(a) and \ref{fig:fig2}(b).
The lowest value of $N_V$ occurs for the same parameters and approaches 0.5.

This behavior is explained as follows. For $\theta = \frac{\pi}{4}$, the exciton is initially prepared in a coherent superposition $\ket{\psi(0)} = \ket{V} = \frac{1}{\sqrt{2}}\qty(\ket{H'} + \ket{V'})$. If the FSS is non-zero, the exciton eigenstates $\ket{H'}$ and $\ket{V'}$ evolve in time with a different phase. As a consequence, the state $\ket{\psi(t)}$ rotates in time between $\ket{V}$ and $\ket{H}$, leading to increased emission into the $H$ cavity mode.
The precession of the exciton state is evident in Fig.~\ref{fig:fig2}(c), where we show the evolution of the state population of the emitter in time for the optimal configuration that maximizes $N_H$.
A similar oscillating behavior can be seen in experiments using polarization-resolved photoluminescence spectroscopy \cite{Schwartz2018, Ollivier2020, Peniakov2025}. 

Two ingredients are necessary to kick-start the precession: (i) the initial state must be a superposition of the exciton eigenstates, and (ii) the exciton eigenvalues must be non-degenerate. This explains why there is no emission into the $H$ mode when $\Delta_{\rm FSS} \approx 0$, and when the rotation angle $\theta$ is an integer multiple of $\frac{\pi}{2}$ (i.e. when the cavity axes are parallel to the exciton axes).
Crucially, the number of $H$ photons is fundamentally bounded to $N_H \leq 0.5$ in this configuration \cite{Ollivier2020}. 

We now introduce an energy splitting $\hbar \Delta_{\rm cav} = $ \SI{770}{\micro\eV} between the cavity modes, which is the value measured in Ref.~\cite{Wang2019a} for a pillar with elliptical cross-section.
In our model, this is implemented by keeping the $H$ mode on resonance with the $H$ exciton, and shifting the $V$ cavity resonance away.
Such a shift suppresses the emission into $V$ significantly and favors the emission into the $H$ mode instead.
As shown in Figs.~\ref{fig:fig2}(d) and \ref{fig:fig2}(e), for $\theta = \frac{\pi}{4}$ and sufficiently large FSS we find $N_H > 0.8$, which is well above the 0.5 threshold observed in the case of the symmetric cavity.
A large FSS is needed so that the precession occurs with a shorter period than the exciton lifetime.
As we explain in the following section, the maximum value of $N_H$ is found in the limit $\Delta_{\rm FSS} \to \infty$ and depends on the ratio between the Purcell-enhanced emission rate in the $H$ mode and the total emission rate. 
Interestingly, we find that the requirement on the rotation angle to maximize $N_H$ is still $\theta = \frac{\pi}{4}$ despite the asymmetry, because this is the only configuration that rotates the initial state $\ket{V}$ fully to $\ket{H}$ after half a precession cycle.
Figure \ref{fig:fig2}(f) shows the evolution of the quantum state of the emitter for $\theta = \frac{\pi}{4}$ and $\hbar \Delta_{\rm FSS} = $ \SI{32.9}{\micro\eV}. We observe that decay to the ground state occurs when the exciton is in $\ket{H}$ but is strongly suppressed when it is in $\ket{V}$, as shown by the step-like behavior of the green curve in Fig.~\ref{fig:fig2}(f). This leads to a larger population of the cavity $H$ mode as compared to the $V$ mode. 

\begin{figure}
    \centering
    \includegraphics[width=\linewidth]{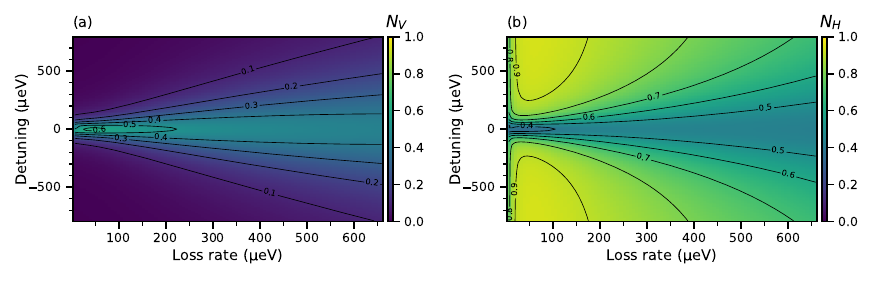}
    \caption{Number of photons emitted with (a) $V$ and (b) $H$ polarization as a function of the cavity loss rate (identical for both polarizations) and of the cavity $V$-mode detuning, $\hbar \delta_V$.
    The $H$ cavity mode is on resonance with the $H$ exciton, i.e.\ $\hbar \delta_H = \frac 1 2 \hbar \Delta_{\rm FSS}$, and the rotation angle is set to $\theta = \pi/4$.
    The FSS is $\hbar \Delta_{\rm FSS} = $ \SI{39.5}{\micro\eV}.
    }
    \label{fig:fig3}
\end{figure}

Next, we investigate the role of cavity parameters on $N_V$ and $N_H$. In Fig.~\ref{fig:fig3}, we fix $\theta=\frac{\pi}{4}$ and $\hbar \Delta_{\rm FSS} = $ \SI{39.5}{\micro\eV} and scan the cavity loss rate $\kappa$ and the mode detuning $\delta_V$, while the $H$ mode is on resonance with the $H$ exciton.
The loss rate is assumed identical for both modes, $\kappa_V = \kappa_H = \kappa$, and determines the bandwidth of the cavity.
By narrowing the cavity bandwidth to $\hbar \kappa = $ \SI{59}{\micro\eV}, we observe a decrease of $N_V$ to approximately zero and an increase of $N_H$ up to a maximum of 0.94 for the largest detuning considered ($\hbar \delta_V = $ \SI{-790}{\micro\eV}).
Once again, the maximum value of $N_H$ increases monotonically with larger $\delta_V$ and can be determined by the appropriate ratio of emission rates in the Purcell-enhanced regime, see next section. 
By further reducing the loss to $\hbar \kappa < $ \SI{50}{\micro\eV}, a decrease in $N_H$ is found. This is due to the transition from weak to strong coupling regime of cavity QED, leading to Rabi oscillations between exciton and cavity populations which interfere with the precession mechanism.

\subsection{Effect of decoherence}
\label{sec:results_decoherence}

\change{
Quantum emitters in a solid state matrix are unavoidably subject to decoherence. The major contributors are charge and spin noise (i.e.\ fluctuating electric and magnetic fields at the location of the emitter, respectively), and coupling to phonons.
In this section, we model decoherence a pure dephasing process occurring at a rate $\gamma$ and examine its effect on the polarized photon output.
To this end, we add two dissipator terms $\gamma \mathcal L_{\dyad{H'}}[\rho]$ and $\gamma \mathcal L_{\dyad{V'}}[\rho]$ to the master equation in Eq.~\eqref{eq:master}.

\begin{figure}
    \centering
    \includegraphics[width=1.0\linewidth]{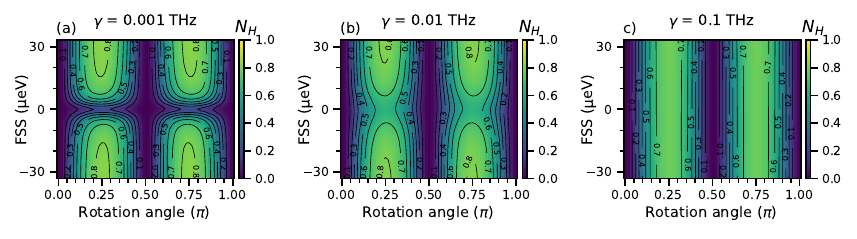}
    \caption{
    \change{Number of photons $N_H$ emitted with $H$ polarization in the presence of pure dephasing at rate (a) $\gamma$ = \SI{0.001}{\tera\Hz}, (b) $\gamma$ = \SI{0.01}{\tera\Hz}, and (c) $\gamma$ = \SI{0.1}{\tera\Hz}. The number $N_H$ is plotted as a function of the rotation angle $\theta$ between cavity and exciton axes and of the FSS $\hbar \Delta_{\rm FSS}$. The splitting between cavity modes is $\hbar \Delta_{\rm cav} = $ \SI{770}{\micro\eV}.}
    }
    \label{fig:figX}
\end{figure}

As shown in Fig.~\ref{fig:figX}, pure dephasing does not change the physics with respect to the rotation angle $\theta$. For any value of $\gamma$ and FSS, the number of photons emitted with $H$ polarization vanishes when $\theta$ is an even multiple of $\frac \pi 4$ and is maximum when $\theta$ is an odd multiple of $\frac \pi 4$.
A moderate value $\gamma$ = \SI{0.001}{\tera\Hz} produces minor differences with respect to the result in the absence of pure dephasing.
For a larger value $\gamma$ = \SI{0.01}{\tera\Hz}, we observe a clear increase in $N_H$ for vanishing FSS, and only a minor decrease of $N_H$ at large FSS. Finally, for $\gamma$ = \SI{0.1}{\tera\Hz}, the result is practically insensitive to the FSS.
To understand this behavior, we notice that the exciton precession time is $T = 2\pi \Delta_{\rm FSS}^{-1}$, which is $T = $ \SI{414}{\pico\second} ($T = $ \SI{138}{\pico\second}) at $\hbar \Delta_{\rm FSS} = $ \SI{10}{\micro\eV} ($\hbar \Delta_{\rm FSS} = $  \SI{30}{\micro\eV}). On the other hand, the value $\gamma$ = \SI{0.1}{\tera\Hz} corresponds to loss of phase coherence on a timescale of \SI{10}{\pico\second}.
When decoherence occurs on a much faster time scale than the exciton precession time, an initial superposition $\ket{\psi(0)} = \ket{V} = \frac{1}{\sqrt{2}}\qty(\ket{H'} + \ket{V'})$ is quickly transformed into an incoherent mixed state instead of precessing.
}

\subsection{Analytical solution in the weak-coupling regime}
\label{sec:results_analytical}

To gain further insight, we now present a simplified model valid in the weak-coupling regime where the cavity modes can be eliminated. In such a regime, the role of the cavity is to enhance the spontaneous emission rate of each exciton state into the respective cavity mode via the well-known Purcell effect.
Then, a model that includes the exciton states with their Purcell-enhanced emission rates $\Gamma_j$, $j \in \qty{H, V}$, is sufficient to reproduce the physics correctly. 
We thus consider the following new Hamiltonian,
\begin{equation}
\label{eq:reduced_model}
    H = \frac{\hbar \Delta_{\rm FSS}}{2} \cos(2\theta) \qty(\dyad{H} - \dyad{V}) + \frac{\hbar \Delta_{\rm FSS}}{2} \sin(2\theta) \qty(\dyad{H}{V} + \dyad{V}{H}) .
\end{equation}
For the master equation, we now use
\begin{align}
\label{eq:reduced_master}
    \dv{t} \rho
    = -\frac{i}{\hbar} \comm{H}{\rho} + \Gamma_H \mathcal L_{\dyad{G}{H}} [\rho] + \Gamma_V \mathcal L_{\dyad{G}{V}} [\rho] + \Lambda \mathcal L_{\dyad{G}{H'}} [\rho] + \Lambda \mathcal L_{\dyad{G}{V'}} [\rho] ,
\end{align}
with rates $\Gamma_j = 4 g_j^2 \kappa_j / \qty(\kappa_j^2 + 4 \delta_j^2)$, \change{as demonstrated rigorously in Appendix \ref{app:appA}}. As done before, we also include decay into undetected background modes at rate $\Lambda$, which for simplicity we assume identical for $H$ and $V$ polarizations. 
Using the column vector $\vb x = \qty(\rho_{HH}, \rho_{VV}, \rho_{HV}, \rho_{VH})^T$, and disregarding other unimportant elements of $\rho$, Eq. \eqref{eq:reduced_master} is equivalent to the following set of differential equations,
\begin{equation}
\label{eq:diff_eq}
    \dv{t} \vb x = 
    \begin{pmatrix}
        -\Gamma_H - \Lambda & 0 & +\frac i 2 \Delta_{\rm FSS} \sin(2\theta) & -\frac i 2 \Delta_{\rm FSS} \sin(2\theta) \\
        0 & -\Gamma_V - \Lambda & -\frac i 2 \Delta_{\rm FSS} \sin(2\theta) & +\frac i 2 \Delta_{\rm FSS} \sin(2\theta) \\
        +\frac i 2 \Delta_{\rm FSS} \sin(2\theta) & -\frac i 2 \Delta_{\rm FSS} \sin(2\theta) & -\frac{\Gamma_H + \Gamma_V}{2} - \Lambda - i \Delta_{\rm FSS} \cos(2\theta) & 0 \\
        -\frac i 2 \Delta_{\rm FSS} \sin(2\theta) & +\frac i 2 \Delta_{\rm FSS} \sin(2\theta) & 0 & -\frac{\Gamma_H + \Gamma_V}{2} - \Lambda + i \Delta_{\rm FSS} \cos(2\theta)
    \end{pmatrix} \vb x .
\end{equation}
The latter can be solved analytically \cite{Mathematica}. From the solution, we calculate the number of photons emitted in each polarization $j$ as $N_j = \Gamma_j \int_0^{+\infty} \dd t \rho_{jj}(t)$.

\subsection{Resonant excitation}

\begin{figure}
    \centering
    \includegraphics[width=1.0\linewidth]{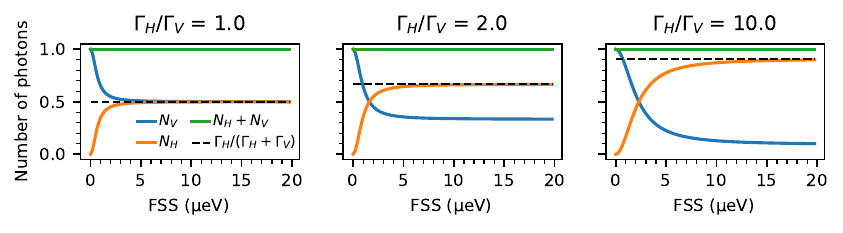}
    \caption{Number of photons emitted in each polarization as a function of the FSS, calculated analytically with Eqs.~\eqref{eq:NV} and ~\eqref{eq:NH}. We use $\hbar \Gamma_V = $ \SI{0.66}{\micro\eV}, corresponding to $(\Gamma_V)^{-1} = $ \SI{1}{\nano\s}, and three increasing values of the ratio $\Gamma_H / \Gamma_V$. The background emission rate $\Lambda$ is set to zero.
    }
    \label{fig:fig4}
\end{figure}

To model the case of resonant excitation in a cross-polarization setup, we set $\rho(0) = \dyad{V}$ as initial condition and we seek to increase the number of photons $N_H$ collected in the $H$ polarization.
For the case of optimal alignment $\theta = \frac \pi 4$, and using $T_j = \Gamma_j + \Lambda$ for the total emission rate from the $j$ exciton, we obtain
\begin{align}
\label{eq:NV}
    N_V & = \frac{\Gamma_V}{T_H + T_V} \frac{T_H^2 + T_H T_V + \Delta_{\rm FSS}^2}{T_H T_V + \Delta_{\rm FSS}^2} ,\\
\label{eq:NH}
    N_H & = \frac{\Gamma_H}{T_H + T_V} \frac{\Delta_{\rm FSS}^2}{T_H T_V + \Delta_{\rm FSS}^2} .
\end{align}
\change{The accuracy of Eqs.~\eqref{eq:NV} and \eqref{eq:NH} is tested in Appendix \ref{app:appA} by comparing them to the results obtained from the numerical solution of the master equation. We find excellent agreement between analytical and numerical results in the relevant weak-coupling regime of cavity QED.}

Let us neglect background emission for the moment (i.e.\ set $\Lambda = 0$ and thus $T_j = \Gamma_j$). As shown in Fig.~\ref{fig:fig4}(a), for identical emission rates $\Gamma_H = \Gamma_V = \Gamma$, we find $N_H = 1 - N_V = \frac 1 2 \frac{\Delta_{\rm FSS}^2}{\Gamma^2 + \Delta_{\rm FSS}^2}$, which implies $N_H \leq 0.5$.
On the other hand, with $\Gamma_H > \Gamma_V$ we find that $N_H$ exceeds the 0.5 threshold as shown in Figs.~\ref{fig:fig4}(b) and \ref{fig:fig4}(c). Specifically, in the limit $\Delta_{\rm FSS}^2 \gg \Gamma_H \Gamma_V$, we find the asymptotic result
$N_j = \Gamma_j / (\Gamma_H + \Gamma_V)$,
showing that the ratio between H- and V-polarized photons is $N_H / N_V = \Gamma_H / \Gamma_V$, i.e.\ it is determined by the ratio of their respective emission rates.
In the same limit, and when background emission is taken into account ($\Lambda > 0$), we obtain
\begin{equation}
\label{eq:large_FSS}
    N_j = \frac{\Gamma_j}{T_H + T_V} .
\end{equation}
Here, the ratio is again $N_H / N_V = \Gamma_H / \Gamma_V$. However, in the presence of significant background emission $N_H$ can decrease below 0.5 even for $\Gamma_H > \Gamma_V$. 
We note that the result $N_j = \Gamma_j / (T_H + T_V)$ has been assumed as the starting point for device optimization in previous work, although without a formal proof \cite{Hoehne2019, Gür2021, Wang2022_nanobeam}. Our result shows that this intuitive formula is valid in the limit $\Delta_{\rm FSS}^2 \gg \Gamma_H \Gamma_V$.

Finally, for $\theta = 0$ we find the obvious result $N_V = \Gamma_V / T_V$ and $N_H = 0$ for any value of the FSS.

\subsection{Above-band excitation}

When indistinguishability of the emitted photons is not required, exciton-based single-photon sources are commonly triggered with above-band excitation.
In this case, the laser excites electrons in the conduction band of the host semiconductor, leaving holes behind. Then, the carriers relax to the QD lowest unoccupied energy state via internal non-radiative processes. This results in random exciton polarization due to the unpredictability of the relaxation process. To model this scenario, we take as initial condition the mixed state $\rho(0) = \frac{1}{2} \qty( \dyad{H} + \dyad{V})$, although inherent asymmetry of the QD can also generate an unbalanced mixture.

Starting with the case $\theta = 0$, i.e.\ when the cavity axes are aligned with the natural exciton axes, the straightforward solution to Eq.~\eqref{eq:diff_eq} is $N_j = \Gamma_j / (2 T_j)$.
Interestingly, in this scenario it is possible to obtain a polarized output (i.e.\ $N_H \ne N_V$) by engineering the emission rates properly; however, the number of emitted photons is still bounded to $N_j \leq 0.50$ for both polarizations.

Moving to $\theta = \frac \pi 4$, the solution to Eq.~\eqref{eq:diff_eq} is now
\begin{align}
    N_V & = \frac{\Gamma_V}{T_H + T_V} \frac{T_H^2 + T_H T_V + 2\Delta_{\rm FSS}^2}{2 T_H T_V + 2 \Delta_{\rm FSS}^2} ,\\
    N_H & = \frac{\Gamma_H}{T_H + T_V} \frac{T_V^2 + T_H T_V + 2\Delta_{\rm FSS}^2}{2 T_H T_V + 2 \Delta_{\rm FSS}^2} .
\end{align}
In the limit of large FSS this leads to the same result $N_j = \Gamma_j / (T_H + T_V)$ as for resonant excitation, which is in principle not bounded to 0.50. Surprisingly, this finding suggests that a rotation of $\theta = \frac \pi 4$ between cavity and exciton axes is also beneficial under above-band excitation, because (for $\Lambda_H \approx \Lambda_V = \Lambda$ and $\Gamma_H > \Gamma_V$)
\begin{equation}
    \frac{N_H(\theta=\pi/4)}{N_H(\theta=0)} = \frac{2 \Gamma_H + 2 \Lambda}{\Gamma_H + \Gamma_V + 2 \Lambda} > 1.
\end{equation}

\section{Charged exciton (trion)}

\begin{figure}
    \centering
    \includegraphics[width=1.0\linewidth]{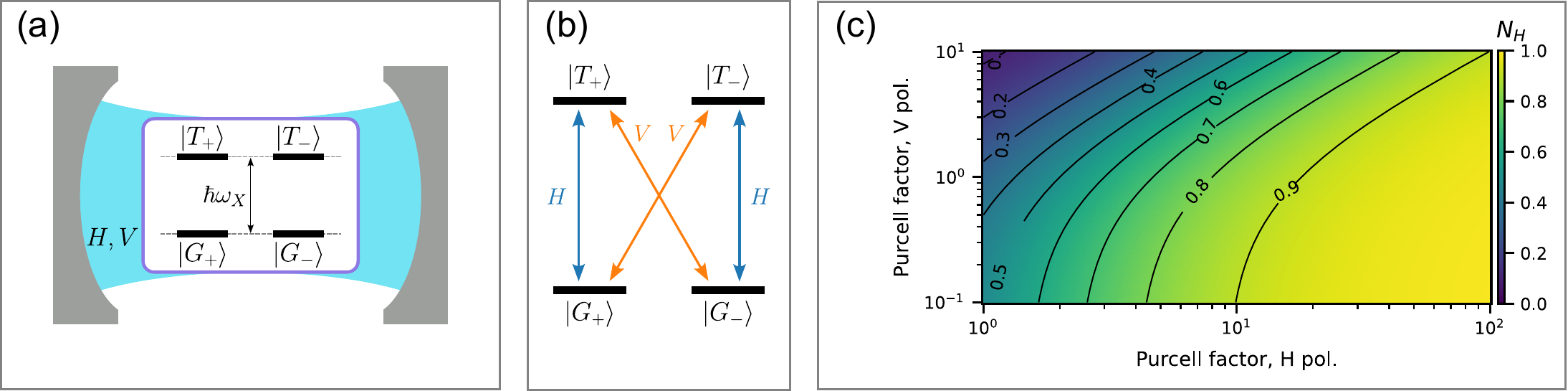}
    \caption{
    (a) Sketch of the energy levels of a charged exciton with eigenstates $\ket{G_\pm}$ and $\ket{X_\pm}$ embedded in a cavity supporting two non-degenerate modes $H$ and $V$.
    (b) Optical selection rules for the charged exciton. Blue (orange) transitions are coupled to $H$-polarized ($V$-polarized) light.
    (c) Number of photons $N_H$ emitted with $H$ polarization as a function of the polarization-dependent Purcell factors, see Eq.~\eqref{eq:N_j_Purcell}. For the background emission factor we use $B = 0.5$.}
    \label{fig:fig5}
\end{figure}

Before moving on to the discussion, we briefly consider the case of an SPS based on a charged exciton (or trion).
Considering, for example, a negatively-charged trion, the emitter is modeled as a four-level system with two ground states $\ket{G_\pm} = \frac{1}{\sqrt{2}} \qty(\ket{\uparrow} \pm \ket{\downarrow})$ and two excited states $\ket{X_\pm} = \frac{1}{\sqrt{2}} \qty(\ket{\uparrow \downarrow \Uparrow} \pm \ket{\downarrow \uparrow \Downarrow})$, where the thin (thick) arrows denote the spin state of the electrons (holes).
This is shown schematically in Fig.~\ref{fig:fig5}(a).
In the absence of magnetic field, the energy levels are two-fold degenerate, i.e.\ $E_{G_{\pm}} = 0$ and $E_{X_\pm} = \hbar \omega_X$. 
The optical selection rules in linear polarization are sketched in Fig.~\ref{fig:fig5}(b).
They entail `vertical' transitions $\ket{X_+} \leftrightarrow \ket{G_+}$ and $\ket{X_-} \leftrightarrow \ket{G_-}$ couple to $H$-polarized light, and `diagonal' transitions $\ket{X_+} \leftrightarrow \ket{G_-}$ and $\ket{X_+} \leftrightarrow \ket{G_-}$ couple to $V$-polarized photons \cite{Ollivier2020, Yao2004, Liu2010}.
In the frame rotating at the exciton frequency and considering different emission rates $\Gamma_H$ and $\Gamma_V$ for $H$- and $V$-polarized emission respectively, the master equation is
\begin{align}
\label{eq:master_eq_trion}
    \dv{t} \rho
    = T_H \qty(\mathcal L_{\dyad{G_+}{X_+}} [\rho] + \mathcal L_{\dyad{G_-}{X_-}} [\rho]) + T_V \qty(\mathcal L_{\dyad{G_+}{X_-}} [\rho] + \mathcal L_{\dyad{G_+}{X_-}} [\rho]) ,
\end{align}
where $T_j = \Gamma_j + \Lambda$, and $\Lambda$ is the background emission rate.
\change{Equation \eqref{eq:master_eq_trion} generates two decoupled differential equations,
\begin{equation}
    \dv{t} \rho_{X_\pm X_\pm} = - \qty(\Gamma_H + \Gamma_V + 2 \Lambda) \rho_{X_\pm X_\pm} ,
\end{equation}
for the diagonal components $\rho_{X_\pm X_\pm}$ of the density operator, which are straightforward to solve.}
Using the definition
\begin{equation}
    N_j = \Gamma_j \int_0^{+\infty} \dd t \Tr \qty[\dyad{X_+} \rho(t) + \dyad{X_-} \rho(t)] ,
\end{equation}
for the number of photons emitted with polarization $j$, we find the result
$N_j = \frac{\Gamma_j}{T_H + T_V}$, valid both in the case of mixed initial state $\rho(0) = \frac{1}{2} \qty(\dyad{X_+} + \dyad{X_-})$ and for pure initial states $\rho(0) = \dyad{X_+}$ or $\rho(0) = \dyad{X_-}$.
This is the same result as for the neutral exciton in the limit $\Delta_{\rm FSS}^2 \gg T_H T_V$, see Eq.~\eqref{eq:large_FSS}.
It is instructive to express it in terms of polarization-dependent Purcell factors $F_j$, $j \in \qty{H, V}$. Using $\Gamma_j = F_j \Gamma_0$, with $\Gamma_0$ the emission rate in a bulk medium, and defining the factor $B = \Lambda / \Gamma_0$ for background emission, we can write
\begin{equation}
\label{eq:N_j_Purcell}
    N_j = \frac{F_j}{F_H + F_V + 2 B} .
\end{equation}
This result is plotted in Fig.~\ref{fig:fig5}(c) as a function of $F_H$ and $F_V$ for a fixed value $B = 0.5$. The figure shows that it is possible to obtain highly-efficient sources ($N_H > 0.9$) even in the presence of modest Purcell enhancement ($F_H \approx 10$), provided that the emission into the opposite mode is strongly suppressed by shifting the $V$ cavity mode, and that the $B$ factor is reduced below 1. The latter is controlled by the lateral dimension of the micropillar \cite{Wang2021_background, Gines2022}.
\change{Pure dephasing has no effect on Eq.~\eqref{eq:N_j_Purcell}, because it affects only the off-diagonal components of $\rho(t)$.}

\section{Discussion}

From our results, we extrapolate guidelines and useful insight for the fabrication of highly-efficient SPS based on a neutral exciton in a micropillar cavity.

The first notable point concerns the optimal alignment between cavity and exciton axes. 
Under resonant excitation, which is often the desired protocol to obtain the highest indistinguishability, it is required to excite the source with a polarized laser and collect the output in the opposite polarization \cite{Somaschi2016}.
The key physical mechanism that makes it possible is the precession of the exciton state.
For a symmetric cavity with degenerate modes, it was shown theoretically in Ref.~\cite{Ollivier2020} that the axes of excitation and collection must form an angle of \SI{45}{\degree} with the exciton axes to maximize the output, and our calculations reproduce this result.
Interestingly, here we demonstrate that a rotation of \SI{45}{\degree} is also required in the case of asymmetric cavities. Thus, the optimal alignment between exciton and cavity axes is not dictated by the symmetry of the problem (circular vs elliptical) but is a fundamental requirement for resonantly-driven SPSs based on a neutral exciton.
During fabrication, we therefore suggest that the device is etched in such a way that the main axes of the elliptical cavity are at \SI{45}{\degree} with respect to the specific crystallographic axes of the host semiconductor, which often dictate the natural exciton polarization.

Second, our results show that a larger splitting $\hbar \Delta_{\rm cav}$ between the cavity modes leads to larger emission in the preferred polarization. This suggests that cavities with very large ellipticity are preferable, because $\Delta_{\rm cav}$ increases with ellipticity. 
However, a larger ellipticity causes an increase in the required laser power---because the emitter is pumped via the misaligned $V$ mode---and a decrease in the collection efficiency---because it reduces the coupling to a Gaussian mode profile in the far field\cite{Wang2019a, Gür2021}.
These two significant drawbacks were not taken into account in our analysis, so we anticipate a trade-off between these competing effects is needed.

Third, we point out that excitons with a large FSS of the order of few tens of \SI{}{\micro\eV} are preferable.
For InAs/GaAs QDs, the natural FSS of the order of \SI{10}{\micro\eV} could limit the  polarized efficiency to sub-optimal values [see e.g. Fig.~\ref{fig:fig2}(e)]. 
However, in the context of entangled photon pair generation, significant efforts have been devoted to control the FSS with electrical and strain fields \cite{Trotta2014, Ollivier2022}, and tuning of the FSS in the \SI{20}{\micro\eV} range has been demonstrated.
The main focus of Refs.~\cite{Trotta2014, Ollivier2022} was to suppress the FSS to improve the entanglement fidelity, but we suggest that similar techniques can be used to enhance the FSS to obtain a highly efficient source of polarized single-photons based on a neutral exciton.
\change{When the exciton loses phase coherence on a time scale $\gamma^{-1} \approx$ 10--100 \SI{}{\pico\second}, it is possible to obtain $N_H > 0.5$ even for vanishing FSS. However, this is detrimental to the indistinguishability, which in this limit is approximated by $I = \Gamma_H (\Gamma_H + \gamma)^{-1}$} \cite{Kiraz2004}.

The result $N_j = \Gamma_j / (T_H + T_V)$, which is valid for the neutral exciton only for $\theta = \frac \pi 4$ and in the asymptotic limit of large FSS, is found to be valid for the charged exciton regardless of the value of $\theta$.
This suggests that the most promising approach for scalable and highly efficient SPSs should rely on trion-based sources. However, the single-photon purity from trion-based sources is generally worse compared to devices based on a neutral exciton \cite{Ollivier2020}, which justifies the interest in boosting the performance of the latter using polarization-selective cavities.

Finally, a comment on the methodology is worth making. For symmetric cavities, a simple and successful approach based on an effective Hamiltonian is available \cite{Ollivier2020}.
There, spontaneous emission is modeled by adding an imaginary component to the exciton frequencies (identical for both polarizations), and the results agree with our numerical and analytical results.
However, we find surprisingly that the effective-Hamiltonian approach leads to ill-defined equations when the imaginary component depends on the exciton polarization, which is needed to consider the case of asymmetric cavities. This is detailed in Appendix \ref{app:appB}.
Therefore, the theoretical formalism reported in this work is necessary to study elliptical cavities.

\section{Conclusions}

We have studied the dynamics of single-photon emission from a QD in a polarization-selective microcavity.
We have presented both a numerical solution of the master equation valid for any set of parameters, and an analytical result which is valid in the weak-coupling regime of cavity QED.
While the polarized photon output of such a source is limited to 0.5 for a symmetric cavity with degenerate modes, our results show that the efficiency can be increased close to unity by splitting the degeneracy, for example using an elliptical shape for the cavity.
For a neutral exciton, it is required that the cavity axes are aligned at \SI{45}{\degree} with respect to the exciton axes, and that the exciton FSS is sufficiently large ($\Delta_{\rm FSS}^2 \gg \Gamma_H \Gamma_V$). On the other hand, these requirements do not apply to the case of a charged exciton.
Our findings demonstrate an avenue towards optimal single-photon sources with simultaneous near-unity efficiency and indistinguishability, which are a fundamental building block for quantum information technologies.

\begin{acknowledgements}

The authors thank Martin A. Jacobsen for stimulating discussions on the modeling of background emission, and for providing the coupling constants and loss rates of the optimized micropillar device.
We acknowledge support from the Novo Nordisk Foundation (grant no. NNF24OC0094739 ``SUPER-Q''), the Independent Research Fund Denmark (grant ID 10.46540/5251-00093B ``INSPEQT''), 
the European Research Council (ERC-CoG ``Unity'', grant no.~865230), the European Union's Horizon 2020 Research and Innovation Programme under the Marie Skłodowska-Curie Grant (Agreement No.~861097), the European Union’s Horizon Europe research and innovation programme under EPIQUE project (Grant Agreement No. 101135288), and the Innovation Fund Denmark (QLIGHT, no. 4356-00002A).
This work was also funded within the QuantERA II Programme that has received funding from the EU H2020 research and innovation programme under GA No 101017733 (via the project ``EQSOTIC'').

\end{acknowledgements}

\appendix

\section{Spontaneous emission rate in the weak coupling regime}
\label{app:appA}

In this Appendix, we prove the formula $\Gamma_j = 4 g_j^2 \kappa_j / \qty(\kappa_j^2 + 4 \delta_j^2)$ that is used in the main text for the Purcell-enhanced emission rate.
Consider a two-level emitter (transition frequency $\omega_X$, raising operator $\sigma^\dag$) coupled to a single cavity mode (mode frequency $\omega_c$, creation operator $a^\dag$). In the frame rotating at frequency $\omega_X$,  the Hamiltonian is
\begin{equation}
    H = \hbar \delta a^\dag a + \hbar g \qty(\sigma^\dag a + \sigma a^\dag) ,
\end{equation}
with $\delta = \omega_c - \omega_X$ the cavity detuning. With the inclusion of cavity leakage at a rate $\kappa$, the master equation reads
\begin{equation}
    \dv{\rho}{t} = -\frac{i}{\hbar} \comm{H}{\rho} + \kappa a \rho a^\dag - \frac{\kappa}{2} \acomm{a^\dag a}{\rho} .
\end{equation}
Representing operators in the basis $\qty{\ket{\alpha} = \ket{X,0}, \ket{\beta} = \ket{G,1}, \ket{\gamma} = \ket{G,0}}$, and focusing only on the relevant elements, we can write the master equation as
\begin{equation}
\label{eq:matrix_form}
    \dv{t}
    \begin{pmatrix} \rho_{\alpha,\alpha} \\ \rho_{\beta,\beta} \\ R_+ \\ R_- \end{pmatrix} = 
    \begin{pmatrix}
    0 & 0 & 0 & ig \\
    0 & -\kappa & 0 & -ig \\
    0 & 0 & -\kappa/2 & i\delta \\
    2ig & -2ig & i\delta & -\kappa/2
    \end{pmatrix}
    \begin{pmatrix} \rho_{\alpha,\alpha} \\ \rho_{\beta,\beta} \\ R_+ \\ R_- \end{pmatrix} ,
\end{equation}
with $R_\pm = \rho_{\alpha,\beta} \pm \rho_{\beta,\alpha}$.
The general solution to \eqref{eq:matrix_form} is $\vb x(t) = \sum_i c_i e^{\lambda_i t} \vb x_i$, with $\lambda_i$ the matrix eigenvalues and $\vb x_i$ the eigenvectors. The eigenvalues are
\begin{widetext}
\begin{align}
    \lambda_{\pm,+} & = -\frac{1}{4} \qty[2 \kappa \pm \sqrt{2} \sqrt{\kappa^2 - 4\delta^2 - 16 g^2 + \sqrt{\qty(\kappa^2 + 4 \delta^2 + 16 g^2)^2 - 64 g^2 \kappa^2} }] , \\
    \lambda_{\pm,-} & = -\frac{1}{4} \qty[2 \kappa \pm \sqrt{2} \sqrt{\kappa^2 - 4\delta^2 - 16 g^2 - \sqrt{\qty(\kappa^2 + 4 \delta^2 + 16 g^2)^2 - 64 g^2 \kappa^2} }] .
\end{align}
In the limit $g \ll \kappa$, we expand the inner square root in powers of $g$ and retain terms up to $g^2$. We find
\begin{align}
   \sqrt{\qty(\kappa^2 + 4 \delta^2 + 16 g^2)^2 - 64 g^2 \kappa^2}
   & = \qty(\kappa^2 + 4 \delta^2) \sqrt{1 - \frac{32 g^2 \qty(\kappa^2 - 4 \delta^2)}{\qty(\kappa^2 + 4 \delta^2)^2} + \frac{16^2 g^4}{\qty(\kappa^2 + 4 \delta^2)^2} } \nonumber \\
   & \approx \qty(\kappa^2 + 4 \delta^2) \qty[1 - 16 g^2 \frac{\kappa^2 - 4 \delta^2}{\qty(\kappa^2 + 4 \delta^2)^2}] .
\end{align}
Inserting into the outer square root, and expanding up to $g^2$, we arrive at the following four eigenvalues, 
\begin{align}
    \lambda_{\pm, +} & = -\frac{\kappa}{2} \qty[1 \pm \sqrt{1 - \frac{16 g^2}{\kappa^2 + 4\delta^2}}] \approx - \frac{\kappa}{2} \qty[1 \pm \qty(1 - \frac{8g^2}{\kappa^2 + 4 \delta^2})] , \\
    \lambda_{\pm, -} & = \qty[-\frac{\kappa}{2} \mp \delta \sqrt{-1 - \frac{16 g^2}{\kappa^2 + 4\delta^2}}] \approx - \frac{\kappa}{2} \mp i \delta \qty(1 + \frac{8g^2}{\kappa^2 + 4 \delta^2}) .
\end{align}
\end{widetext}
All four eigenvalues have a negative real part, resulting exponential suppression of the exciton population $\rho_{\alpha,\alpha}(t)$ in time. 
The eigenvalue with the smallest absolute value of the real part, which represents the slowest decay and thus dominates the dynamics in the asymptotic limit, is $\lambda_{+,-} = -\frac{4 g^2 \kappa}{\kappa^2 + 4 \delta^2}$. 
It follows that the exciton population decays approximately as $P_X(t) \propto e^{-\Gamma t}$, with 
\begin{equation}
\label{eq:emission_rate}
    \Gamma = \frac{4 g^2}{\kappa} \frac{\kappa^2}{\kappa^2 + 4 \delta^2}.
\end{equation}
This formula is used for the cavity-enhanced emission rates in Sec.~\ref{sec:results_analytical}.

\begin{figure}
    \centering
    \includegraphics[width=\linewidth]{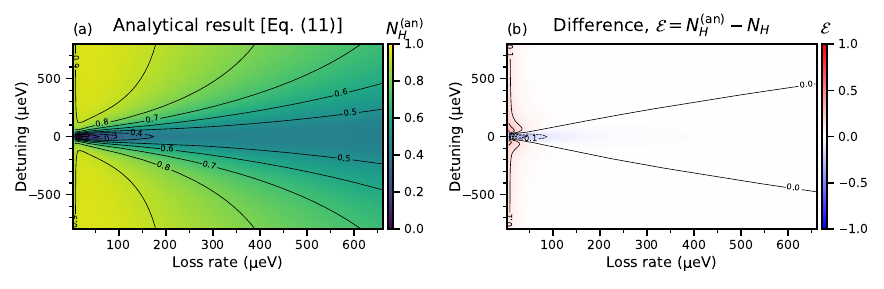}
    \caption{
    (a) Number of photons emitted with $H$ polarization as a function of the cavity loss rate (identical for both polarizations) and of the cavity $V$-mode detuning, $\hbar \delta_V$, calculated analytically using Eq.~\eqref{eq:NH}.
    The polarization-dependent emission rates used in Eq.~\eqref{eq:NH} are calculated from the detuning and the loss rate using Eq.~\eqref{eq:emission_rate}. Other parameters are as in Fig.~\ref{fig:fig3}.
    (b) Difference between the analytical result $N_H^{\rm (an)}$ of panel (a) and the numerical result $N_H$ shown in Fig.~\ref{fig:fig3}(b).}
    \label{fig:fig6}
\end{figure}

In Fig.~\ref{fig:fig6}(a) we calculate the number of photons emitted with $H$ polarization using the analytical result in Eq.~\eqref{eq:NH} and with emission rates calculated from Eq.~\eqref{eq:emission_rate}. We use the same parameters as in Fig.~\ref{fig:fig3}(b), where the data points were calculated with the numerical master-equation approach.
The two results are in excellent agreement. The difference between the numerical and analytical results is shown in Fig.~\ref{fig:fig6}(b).

\section{Breakdown of the effective non-Hermitian Hamiltonian for polarization-dependent emission rate.}
\label{app:appB}

In this Appendix, we calculate the emitted photon numbers $N_V$ and $N_H$ using an effective-Hamiltonian approach.
We model spontaneous emission by adding an imaginary component $-i \frac{\Gamma_j}{2}$ to the exciton frequencies $\omega_j$, $j \in \qty{H, V}$, leading to the following non-Hermitian Hamiltonian \cite{Ollivier2020},
\begin{equation}
    \label{eq:eff_H}
    H = \hbar \qty(\omega_H - i \frac{\Gamma_H}{2}) \dyad{H'} + \hbar \qty(\omega_V - i \frac{\Gamma_V}{2}) \dyad{V'} .
\end{equation}
As we demonstrate below, this approach is valid when $\Gamma_H = \Gamma_V$ but it breaks down when $\Gamma_H \ne \Gamma_V$

First, we initialize the system in the superposition state $\ket{\psi(0)} = \ket{V} = \cos(\theta) \ket{V'} + \sin(\theta) \ket{H'}$. The time evolution $\ket{\psi(t)}$ is straightforward, and its projection onto $\ket{V}$ and $\ket{H}$ is 
\begin{align}
    \label{eq:pV}
    \braket{V}{\psi(t)} & = \cos^2(\theta) e^{-i \qty(\omega_V + \frac{\Gamma_V}{2}) t} + \sin^2(\theta) e^{-i \qty(\omega_H + \frac{\Gamma_H}{2})t} , \\
    \label{eq:pH}
    \braket{H}{\psi(t)} & = \frac{\sin(2\theta)}{2} \qty[e^{-i \qty(\omega_H + \frac{\Gamma_H}{2})t} - e^{-i \qty(\omega_V + \frac{\Gamma_V}{2})}] .
\end{align}
Observing that the number $\dd N_j$ of photons emitted with polarization $j$ in the time interval between $t$ and $t + \dd t$ is $\dd N_j = \Gamma_j p_j(t) \dd t$, with $p_j (t) = \qty|\braket{j}{\psi(t)}|^2$, we define $N_V$ and $N_H$ as
\begin{equation}
    N_j = \Gamma_j \int_0^{+\infty} \dd t \qty|\braket{j}{\psi(t)}|^2 .
\end{equation}
Substituting Eqs.~\eqref{eq:pV} and \eqref{eq:pH} we obtain
\begin{align}
    \label{eq:NV_effH}
    N_V & = \cos^4(\theta) + \frac{\Gamma_V}{\Gamma_H} \sin^4(\theta) + \frac{\sin^2(2 \theta) \qty(\Gamma_H + \Gamma_V) \Gamma_V}{\qty(\Gamma_H + \Gamma_V)^2 + 4 \Delta_{\rm FSS}^2} , \\
    \label{eq:NH_effH}
    N_H & = \sin^2(2 \theta) \frac{\Gamma_H + \Gamma_V}{4 \Gamma_V} \frac{\qty(\Gamma_H -\Gamma_V)^2 + 4 \Delta_{\rm FSS}^2}{\qty(\Gamma_H + \Gamma_V)^2 + 4 \Delta_{\rm FSS}^2} .
\end{align}
For identical emission rate $\Gamma_H = \Gamma_V = \Gamma$, we obtain
\begin{align}
    N_H = \frac{\sin^2(2 \theta)}{2} \frac{\Delta_{\rm FSS}^2}{\Gamma^2 + \Delta_{\rm FSS}^2}
\end{align}
and $N_V = 1 - N_H$, which is identical to the result of Ref.~\cite{Ollivier2020}. In particular, this shows that $N_H$ is maximum at $\theta = \frac{\pi}{4}$ but is bounded to $N_H \leq \frac 1 2$. 
However, for $\Gamma_H \ne \Gamma_V$ we observe that the sum $N_V + N_H$ is not normalized to 1. For example, for $\theta = \frac{\pi}{4}$ and in the limit $\Delta_{\rm FSS} \gg \Gamma_H, \Gamma_V$ we obtain $N_H + N_V = \frac{\qty(\Gamma_H + \Gamma_V)^2}{4 \Gamma_H \Gamma_V}$, which is always $>1$ except for $\Gamma_H = \Gamma_V$.
This is a flaw of the effective Hamiltonian approach and makes it impossible to apply such a simplified model in our work.

To understand the origin of this issue, and explain why the special case $\Gamma_H = \Gamma_V$ is not affected, we point out that the state $\ket{\psi(t)}$ is not normalized,
\begin{equation}
    \braket{\psi(t)} = \cos^2(\theta) e^{-\Gamma_V t} + \sin^2(\theta) e^{-\Gamma_H t} \ne 1.
\end{equation}
This is a consequence of the fact that the Hamiltonian in Eq.~\eqref{eq:eff_H} generates a non-unitary dynamics.
In the special case $\Gamma_H = \Gamma_V = \Gamma$, the inner product is $\braket{\psi(t)} = e^{-\Gamma t}$ and satisfies $\int_0^{+\infty} \dd t \braket{\psi(t)} = \Gamma^{-1}$. It follows that
\begin{align}
    N_V + N_H
    & = \Gamma \int_0^{+\infty} \dd t \qty[ \qty|\braket{V}{\psi(t)}|^2 + \qty|\braket{H}{\psi(t)}|^2]  = \frac{\int_0^{+\infty} \dd t \braket{\psi(t)}}{\int_0^{+\infty} \dd t \braket{\psi(t)}} = 1 ,
\end{align}
which guarantees that the total number of emitted photons is normalized to 1.
This does not hold for $\Gamma_H \ne \Gamma_V$, due to the fact that $\int_0^{+\infty} \dd t \braket{\psi(t)} = \Gamma_V^{-1} \cos^2(\theta) + \Gamma_H^{-1} \sin^2(\theta)$.

In an attempt to avoid this issue, one can enforce the correct normalization by defining
\begin{equation}
    \widetilde N_j = \frac{\int_0^{+\infty} \dd t \qty|\braket{j}{\psi(t)}|^2}{\int_0^{+\infty} \dd t \braket{\psi(t)}} .
\end{equation}
However, this leads to the unphysical result that $N_H < \frac 1 2$ for any choice of the parameters. Another strategy is to define
\begin{equation}
    \widehat N_j = \frac{\Gamma_j \int_0^{+\infty} \dd t \qty|\braket{j}{\psi(t)}|^2}{\sum_j \Gamma_j \int_0^{+\infty} \dd t \qty|\braket{j}{\psi(t)}|^2} ,
\end{equation}
however, this makes it possible to obtain $N_H > N_V$ in the absence of FSS, which is also incorrect.

\bibliography{biblio}

\end{document}